\documentclass[a4paper]{article}

\usepackage{amsfonts}

\def\be{\begin{equation}}
\def\ee{\end{equation}}
\def\bea{\begin{eqnarray}}
\def\eea{\end{eqnarray}}
\def\({\left(}
\def\){\right)}
\def\<{\left<}
\def\>{\right>}

\def\[{\left[}
\def\]{\right]}
\def\tr{{\mbox{tr}}}
\def\be{\begin{equation}}
\def\ee{\end{equation}}
\def\bea{\begin{eqnarray}}
\def\eea{\end{eqnarray}}
\def\({\left(}
\def\){\right)}
\def\<{\big<}
\def\>{\big>}

\def\[{\left[}
\def\]{\right]}
\def\+{\bar}

\def\tr{{\mbox{tr}}}
\def\a{{\cal{A}}}
\def\b{{\cal{B}}}

\begin{document}

\pagestyle{empty}
\vskip-10pt
\vskip-10pt
\hfill 
\begin{center}
\vskip 3truecm
{\Large \bf
Algebraic structures on parallel M2-branes}\\ 
\vskip 2truecm
{\large \bf
Andreas Gustavsson}\footnote{a.r.gustavsson@swipnet.se}\\
\vskip 1truecm
{\it F\"{o}rstamajgatan 24,\\
S-415 10 
G\"{o}teborg, Sweden}\\
\end{center}
\vskip 2truecm
{\abstract{In the course of closing supersymmetry on parallel M2 branes up to a gauge transformation, following the suggestion in hep-th/0611108 of incorporating a gauge field which only has topological degrees of freedom, we are led to assume a certain algebraic structure for the low energy theory supposedly living on parallel M2 branes.}}

\vfill 
\vskip4pt
\eject
\pagestyle{plain}

\section{Introduction}
The low-energy theory living on a single M2 brane was derived in \cite{Kaplan:1995cp}, \cite{Adawi:1998ta}. This theory can also be derived from the Yang-Mills action living on a single D2 brane by dualizing one of the scalar fields \cite{Bergshoeff:1996tu}. 

In this Letter we investigate the `non-abelian' generalization of this M2 brane theory, much inspired by the work of Bagger and Lambert \cite{Bagger:2006sk}. More specifically, we ask what requirements come from supersymmetry. Our starting point is to assume that supersymmetry can be realized on some kind of `fields' with the usual Lorentz index structure. We do not need to know much about the internal structure to be able to analyze the supersymmtry transformations. All we in essence need, is usual gamma matrix algebra.

We put non-abelian in quotion marks because the fields will not take values in the adjoint{\footnote{Since the gauge field would be in the adjoint, all other fields would also be in the adjoint due to supersymmetry.} representation of a non-abelian Lie algebra. We also put the word `fields' in quotion marks because it is not clear that these would be just ordinary fields. Perhaps they are `non-abelian loops'. 

We thus assume there are some kind of non-commuting `fields' that take values in some algebra, and that there are certain ways of multiplying together such fields to get a new element in the algebra. We also require that these fields are such that they reduce to the ordinary eight scalar fields plus their supersymmetric partners in the abelian case.

Dimensional analysis suggests there is in the supersymmetry variations, products of two as well as of three fields. It is of course desirable that all products of our fields are such that they close on some internal algebra. The way we do that is by making the minimal assumption that there is a multiplication of two fields which belong to some set of fields, that we denote as $\a$, such that the product is in some set $\b$ that need not be the same as $\a$, though a product of three elements in $\a$ must yield back an element in $\a$. We introduce three different kinds of multiplications to be used according to whether the elements being multiplied belong to $\a$ or $\b$. Then we see what requirements closure of the supersymmetry transformations impose on these various multiplications. 

We end this Letter by proposing a rather explicit, though maybe drastic, way of realizing these multiplication operations. For this construction we abandon the concept of ordinary fields and instead assume the fields are loops in transverse space.

\section{The abelian cases}
The abelian super Yang-Mills SUSY transformations can be derived by dimensionally reducing $1+9$ dimensional super Yang-Mills to $1+2$ dimensions. The ten-dimensional spinor is Majorana-Weyl,
\bea
\Gamma^{(10)}\chi = \chi
\eea
Since we wish to prepare the ground for an up-lift to M-theory, it is desirable to look for an embedding of $SO(1,9)$ into $SO(1,10)$ in which $\Gamma^{(10)}$ is the eleveth gamma matrix. We denote the gamma matrices as $\Gamma^M$ ($M=0,...9,10$). In M-theory, the spinor can obviously just be Majorana. But the presence of an $M2$ brane breaks the Lorentz symmetry as $SO(1,10) \rightarrow SO(1,2)\times SO(8)$, and we can have a Weyl spinor of $SO(8)$. Let us denote by 
\bea
\Gamma = \Gamma^{3..9(10)}
\eea
the chirality matrix of $SO(8)$. Half the supersymmetry is broken by the M2 brane. Let us choose a convention where the unbroken supersymmetry parameters are Weyl
\bea
\Gamma \epsilon = \epsilon
\eea
The broken supersymmetries gets manifested as goldstinos,
\bea
\Gamma \psi = -\psi
\eea
living on the world-volume of the M2 brane. The broken translations become eight goldstone scalar fields, $\phi^A$ on the M2 brane. The unbroken supersymmetries relate the bosonic and fermions degrees of freedom. The Yang-Mills supersymmetry transformations, written in terms of such spinors, is given by \cite{Bagger:2006sk}
\bea
\delta \phi^a &=& i\bar{\epsilon}\Gamma^a\psi\cr
\delta F_{\mu\nu} &=& -2i\bar{\epsilon}\Gamma_{[\mu}\Gamma^{(10)}\partial_{\nu]}\psi\cr
\delta \psi &=& \frac{1}{2}F_{\mu\nu}\Gamma^{\mu\nu}\Gamma^{(10)}\epsilon + \partial_{\mu}\phi^a\Gamma^{\mu}\Gamma_a\epsilon
\eea

Let us now dualize the gauge field to a scalar field \footnote{This corresponds to equating $\phi^{(10)}$ with the dual gauge potential. The intergral of a dual potential over a closed loop is well-defined only modulo $2\pi$, and since we are in three dimensions, the dual potential is a zero-form and integration of a zero-form should mean evaluation of that zero form at a point, suggesting this scalar would be compact,
\bea
\phi^{(10)} \sim \phi^{(10)} + 2\pi R.
\eea}
\bea
\partial_{\mu}\phi^{(10)} = \frac{R}{2}\epsilon_{\mu\nu\rho}F^{\nu\rho}
\eea

We then get (if we also replace $\phi^a$ by the reduced field $R\phi^a$ and $\psi$ by the reduced field $R\psi$ respectively)
\bea
\delta \psi = \partial_{\mu}\phi^{(10)}\Gamma^{\mu}\Gamma_{(10)}\epsilon + \partial_{\mu}\phi^a\Gamma^{\mu}\Gamma_a\epsilon
\eea
and
\bea
\delta \partial_{\mu}\phi^{(10)} &=& 2i\bar{\epsilon}\Gamma_{\mu\nu}\Gamma^{(10)}\partial^{\nu}\psi 
\eea
To proceed we must use the equation of motion
\bea
\Gamma^{\nu}\partial_{\nu}\psi=0
\eea
Then we can use $\Gamma_{\mu\nu} = -\eta_{\mu\nu} + \Gamma_{\mu}\Gamma_{\nu}$, and write this as
\bea
\delta \partial_{\mu}\phi^{(10)} &=& 2i\bar{\epsilon}\Gamma^{(10)}\partial_{\mu}\psi 
\eea

This way we have reached the $SO(8)$ covariant SUSY transformations of an M2 brane,
\bea
\delta \phi^A &=& i\bar{\epsilon} \Gamma^A \psi\cr
\delta \psi &=& \partial_{\mu}\phi^A \Gamma^{\mu}\Gamma_A \epsilon
\eea

In conformal four-dimensional Yang-Mills theory we have the length dimensions
\bea
(A_{\mu},\phi^a,\psi,\epsilon) = \(-1,-1,-\frac{3}{2},\frac{1}{2}\)
\eea
Reduction to three dimensions, does not change these dimensions. But we rather get a dimensionful coupling constant. 

The theory living on an M2 brane is conformal. So the fields acquire the length dimensions
\bea
(A_{\mu},\phi^A,\psi,\epsilon) = \(-1,-\frac{1}{2},-1,\frac{1}{2}\).
\eea
Intuitively one may perhaps think of this as that the Goldstone modes get their length-dimensions increased by $\frac{1}{2}$ as a result of an integration over one compact transverse dimension.\footnote{The Einstein metric (or the M-theory metric) $G$ is related to the string metric $g$ by the Weyl rescaling $g=G/R$ where $R$ is the radius of the M-theory circle as measured by the Einstein metric. In the string metric a transverse dimension has length dimension $1$, and scalars on the D-brane are rescaled by one power of $\alpha'$ so as to get dimension $-1$. In Einstein metric transverse dimension gets dimension $1/2$ and scalars on the brane get dimension $-1/2$.} But the gauge field does not. It will become even more apparent that the gauge field is to be treated very differently from the scalars and fermions when we turn to the non-abelian case. As we will see, the gauge field will be much like an ordinary field, taking values in an algebra $\b$ that closes on itself, while the scalars and fermions will belong to a different set $\a$ that is like the square root of $\b$.

\section{Non-abelian generalization}
We will assume the algebra is a semi-direct product of two distinct sets, that we denote as $\a$ and $\b$ respectively. We introduce three different kinds of products,
\bea
\<\cdot,\cdot\> &:&\a\times \a \rightarrow  \b\cr
(\cdot,\cdot) &:&\a\times \b \rightarrow  \a\cr
[\cdot,\cdot] &:&\b\times \b \rightarrow  \b
\eea
We get the familiar algebraic structure of Yang-Mills theory if we assume that $\a=\b$ is an ordinary Lie algebra, and that all these multiplications are identical.

We will assume the products have the following properties
\bea
\<\alpha,\beta\> &=& -\<\beta,\alpha\>\cr
(A,a) &=& \varepsilon(a,A)\label{q1}\cr
[A,B] &=& -[B,A],\label{eq1}
\eea
the `associative' property
\bea
(\alpha,\<\beta,\gamma\>) = \varepsilon(\<\alpha,\beta\>,\gamma) \label{cycl}
\eea
though we let $\varepsilon$ be an arbitrary sign factor for the time being. We also assume these products are subject to the following Jacobi identities,
\bea
\<(A,\alpha),\beta\> - \<(A,\beta),\alpha\> &=& [A,\<\alpha,\beta\>]\label{jac1}\\ 
((\alpha,A),B) - ((\alpha,B),A) &=& -\varepsilon(\alpha,[A,B])\\ \label{jac2}
[[A,B],C] - [[A,C],B] &=& [A,[B,C]]\label{jac3}.
\eea
These will, among other things, be crucial when verifying that the gauge transformations close into a gauge algebra, and when verifying equations of motion are gauge invariant.
 
It is of course not obvious these assumptions are consistent with each other. Either one may try and look for inconsistencies in any of the Jacobi identities and the other assumptions. But there appears to be a simpler way of checking consistency of these assumptions. Namely by noting there is an explicit, finite-dimensional realization of this algebra. We then take as $\a$ the set of gamma matrices $\gamma^i$ in four dimensions (as suggested by scalar fields taking values in a fuzzy three sphere, \cite{Basu:2004ed}), and as $\b$ the set of $\gamma^{ij}$. We then define the products as $\<\gamma^i,\gamma^j\> = [\gamma^i,\gamma^j]\gamma_5 \in \b$, $(\gamma^{jk},\gamma^i)\equiv \varepsilon(\gamma^i,\gamma^{jk}) = -[\gamma^i,\gamma^{jk}] \in \a$ and $[\gamma^{ij},\gamma^{kl}] = [\gamma^{ij},\gamma^{kl}] \in \b$. In the right-hand sides, the brackets denote usual commutators.\footnote{We notice the presence of the $\gamma_5$-factor in $\<\cdot,\cdot\>$. This will correspond to the matrix $G_5$ that appears in the Basu-Harvey equation \cite{Basu:2004ed} as our Basu-Harvey equation will be of the form $dT^i/ds \sim \epsilon_{ijkl}(T^j,\<T^k,T^l\>)$.} We note that  $[\cdot,\cdot]:\b\rightarrow \b$ follow as a consequence of the Lorentz algebra. One may also verify that the associativity property and the Jacobi identities are satisfied. It is clear though, that this realization, nor any fuzzy three-sphere generalization of the gamma-matrices, can reasonably contain all information of the gauge group that is supposed to appear when we reduce to D2 branes. So we should look for a larger realization of this algebra. 

We introduce a gauge covariant derivative $D_{\mu}=\partial_{\mu}+A_{\mu}$ as usual. Acting with this derivative on an element in either $\a$ or $\b$, we should get back an element in the same set. Taking two derivatives we should still end up with an element in the same set as we started with. Hence $D_{\mu}$ should be an operator in $\b$. In particular we must take $A_{\mu}\in \b$. 

Gauge variations act as
\bea
\delta_{\Lambda}A_{\mu} &=& -\varepsilon D_{\mu}\Lambda\cr
\delta_{\Lambda}\phi^A &=& (\phi^A,\Lambda)\cr
\delta_{\Lambda}\psi &=& (\psi,\Lambda)
\eea
and they form a closed gauge algebra. To see that, we compute
\bea
[\delta_{\Lambda'},\delta_{\Lambda}]A_{\mu} &=& -\varepsilon \([\delta_{\Lambda'}A_{\mu} ,\Lambda]-[\delta_{\Lambda}A_{\mu},\Lambda']\)\cr
&=& [D_{\mu}\Lambda',\Lambda]-[D_{\mu}\Lambda,\Lambda'] = D_{\mu}[\Lambda',\Lambda],
\eea
and
\bea
[\delta_{\Lambda'},\delta_{\Lambda}]\phi &=& ((\phi,\Lambda'),\Lambda)-((\phi,\Lambda),\Lambda')\cr
&=& -\varepsilon (\phi,[\Lambda',\Lambda])
\eea
from which we deduce the gauge algebra
\bea
[\delta_{\Lambda'},\delta_{\Lambda}] = -\varepsilon \delta_{[\Lambda',\Lambda]}.
\eea

The strange sign factor in the gauge variation is necessary in order for $D_{\mu}\phi$ to transform covariantly,
\bea
\delta_{\Lambda}(D_{\mu}\phi) &=& (\delta_{\Lambda}A_{\mu},\phi) + D_{\mu}\delta_{\Lambda}\phi\cr
&=& -\varepsilon (D_{\mu}\Lambda,\phi) + D_{\mu}(\phi,\Lambda)\cr
&=& (D_{\mu}\phi,\Lambda).
\eea

Just as in the abelian case, we would like to dualize the Yang-Mills gauge field into a scalar field to get the theory on M2 branes. However this is unlikely the whole story in the non-abelian case -- it seems not possile to derive the M2 brane theory this way. When we dualize the Yang-Mills gauge field it should mean it does no longer enter our equations after dualization. That suggests we should dualize the gauge field as follows,
\bea
D_{\mu}^{NEW}\phi^{(10)} = \frac{R}{2}\epsilon_{\mu\nu\rho}F^{\nu\rho}_{OLD}
\eea
We do not want the Yang-Mills gauge field to occur at anywhere after dualization, so we have introduced a new gauge field that enters in the covariant derivative on the left-hand side.

In the sequel we drop the subscript $NEW$, it always being understood that we use the new flat gauge field. It seems very unlikely we get the M2 theory by dualizing the Yang-Mills gauge field like this, but nevertheless we find this dualization useful. 

The most general ansatz for the supersymmetry transfomations, consistent with $SO(1,2)\times SO(8)$ symmetry and dimensional analysis, appears to be
\bea
\delta \phi^A &=& i\bar{\epsilon}\Gamma^A\psi\cr
\delta A_{\mu} &=& i\alpha\bar{\epsilon}\Gamma_{\mu}\Gamma^A\<\psi,\phi_A\>\cr
\delta \psi &=& \gamma\Gamma^{\mu}\Gamma_A\epsilon D_{\mu}\phi^A + \beta \Gamma_{ABC}\epsilon (\phi^A,\<\phi^B,\phi^C\>) 
\eea
for some numerical coefficients $\alpha$, $\beta$, $\gamma$, to be determined. 

It is now clear why we required the Eqs (\ref{q1}), (\ref{cycl}). We then find that $(\phi^A,\<\phi^B,\phi^C\>) = \varepsilon^2(\phi^B,\<\phi^C,\phi^A\>)$ and  
\bea
(\phi^A,\<\phi^B,\phi^C\>) &=& \varepsilon(\<\phi^A,\phi^B\>,\phi^C)\cr
& =& -\varepsilon(\<\phi^B,\phi^A\>,\phi^C)\cr
& =& -\varepsilon^2(\phi^B,\<\phi^A,\phi^C\>),
\eea
We now see that we can get the symmetries of $\Gamma_{ABC}$ if and only if $\varepsilon^2 = 1$. 

We notice that the field strength is absent in the variation of the fermions. If we assume the action for the gauge field being the Chern-Simons action \cite{Schwarz:2004yj}, then we should get source terms which are bilinears in the fermionic field when varying the gauge field in the gauge covariant derivatives. So the gauge field strength need not vanish, despite it contains no local physical degrees of freedom. 

There appears to be no way of incorporating the field strength in $\delta \psi$ consistent with dimensional analysis.

As in the abelian case, the term
\bea
D_{\mu}\phi^{(10)}\Gamma_{(10)}\epsilon
\eea
arose from dualizing the corresponding term
\bea
F_{\mu\nu}\Gamma^{\mu\nu}\Gamma^{(10)}\epsilon
\eea
in Yang-Mills theory. If we take the viewpoint that the gauge field is new, its transformation rule can not be derived from Yang-Mills theory. But dimensional analyzis suggests this form. If we take the viewpoint that it is the flat piece of the Yang-Mills conection, it becomes suggestive that it does indeed reduce to the transformation rule of a gauge field if we compactify $\phi^{(10)}$.

Let us now turn to the issue of closure of supersymmetry.

\subsection{The scalars}
We begin by the scalars. We have 
\bea
[\delta_{\eta},\delta_{\epsilon}]\phi^A &=& -2i\gamma\bar{\epsilon}\Gamma^{\mu}\eta D_{\mu}\phi^A\cr
&&+ 6\beta i\bar{\epsilon}\Gamma_{BC}\eta (\phi^{[A},\<\phi^B,\phi^{C]}\>)
\eea
Since we have assumed that our brackets have a cyclic property, we can write the result as
\bea
[\delta_{\eta},\delta_{\epsilon}]\phi^A &=& -2i\gamma\bar{\epsilon}\Gamma^{\mu}\eta D_{\mu}\phi^A + (\phi^{A},\Lambda)
\eea
Here  
\bea
\Lambda = 6i\beta\bar{\epsilon}\Gamma_{AB}\eta \<\phi^A,\phi^B\>\label{gauge1}
\eea
should be given the interpretation of a gauge parameter.

\subsection{The gauge field}
We compute
\bea
\alpha^{-1}\delta_{\eta}\delta_{\epsilon}A_{\mu} &=& i\bar{\epsilon}\Gamma_{\mu}\Gamma_A\<\delta_{\eta}\psi,\phi^A\>+i\bar{\epsilon}\Gamma_{\mu}\Gamma_A\<\psi,\delta_{\eta}\phi^A\>\cr
&=& i\gamma\bar{\epsilon}\Gamma_{\mu}\Gamma_A \Gamma^{\nu}\Gamma_B \eta \<D_{\nu} \phi^B, \phi^A\>\cr
&& + i\beta\bar{\epsilon}\Gamma_{\mu}\Gamma_A\Gamma_{BCD}\eta \<(\phi^B,\<\phi^C,\phi^D\>),\phi^A\>\cr
&& - \bar{\epsilon}\Gamma_{\mu}\Gamma_A\psi \bar{\eta}\Gamma^A\psi
\eea
We note that, by applying the Jacobi identity,
\bea
&&((\phi^{[A},\<\phi^B,\phi^C\>),\phi^{D]})\cr
&=&\varepsilon((\<\phi^{[A},\phi^B\>,\phi^C),\phi^{D]})\cr 
&=&\varepsilon[\<\phi^{[A},\phi^B\>,\<\phi^C,\phi^{D]}\>] \equiv 0.
\eea
Applying a Fierz rearrangement on the last term, we now get (on-shell)
\bea
[\delta_{\eta},\delta_{\epsilon}]A_{\mu} &=& -2i\gamma(\bar{\epsilon}\Gamma^{\kappa}\eta)F_{\kappa\mu} + D_{\mu}\Lambda
\eea
We thus assume the equation of motion 
\bea
F_{\mu\nu} = \alpha \epsilon_{\mu\nu\rho}\<D^{\rho}\phi^A,\phi_A\> + \frac{i\alpha}{2}\<\bar{\psi},\Gamma_{\mu\nu}\psi\>\label{constraint}
\eea
and the gauge parameter
\bea
\Lambda = i\alpha\gamma\bar{\epsilon}\Gamma_{AB}\eta \<\phi^A,\phi^B\>.\label{gauge2}
\eea

\subsection{The fermions}
We now turn to the fermions:
\bea
[\delta_{\eta},\delta_{\epsilon}]\psi &=& i\gamma\Gamma^{\mu}\Gamma_A (\epsilon\bar{\eta}-\eta\bar{\epsilon}) \Gamma^A D_{\mu}\psi\cr
&&+i\alpha\Gamma^{\mu}\Gamma_A (\epsilon\bar{\eta}-\eta\bar{\epsilon}) \Gamma_{\mu}\Gamma_B(\<\psi,\phi^B\>,\phi^A) \cr
&&+3i\beta \Gamma_{ABC} (\epsilon\bar{\eta}-\eta\bar{\epsilon})\Gamma^A(\psi,\<\phi^B,\phi^C\>)
\eea
We now use Eq (\ref{cycl}) and a Fierz rearrangement, derived in the appendix. We then get\footnote{I am grateful to Ulf Gran for checking the various gamma matrix identities in Eq (\ref{gammas}) using the computer program GAMMA \cite{Gran:2001yh}, and for actually localizing the term in the second of these identities that was missing in previos versions of this paper. This missing term is the reason I failed to show on-shell closure in previous versions of this paper.} 
 \bea
[\delta_{\eta},\delta_{\epsilon}]\psi &=& -2i\gamma(\bar{\epsilon}\Gamma^{\kappa}\eta)D_{\kappa}\psi\cr
&&+i(\bar{\epsilon}\Gamma_{\kappa}\eta)\Gamma^{\kappa}\(\gamma\Gamma^{\mu}D_{\mu}\psi - \(\frac{9}{4}\beta-\frac{1}{8}\alpha\varepsilon\)\Gamma_{AB}(\psi,\<\phi^A,\phi^B\>\)\cr
&&- \frac{1}{4}i(\bar{\epsilon}\Gamma_{EF}\eta)\Gamma^{EF} \(\gamma\Gamma^{\mu} D_{\mu} \psi + \(\frac{3}{2}\beta+\frac{3}{4}\alpha\varepsilon\)\Gamma_{BC}(\psi,\<\phi^B,\phi^C\>\)\cr
&&-\frac{1}{16}i(-24\beta+12\alpha\varepsilon)(\bar{\epsilon}\Gamma_{EF}\eta)\delta^E_B\Gamma^F\Gamma_C(\psi,\<\phi^B,\phi^C\>)\cr
&&-6i\beta(\bar{\epsilon}\Gamma_{EF}\eta)\Gamma_B{}^F(\psi,\<\phi^E,\phi^B\>)\cr
&&-\frac{6\beta+\alpha\varepsilon}{16.24}i(\bar{\epsilon} \Gamma_{\kappa}\Gamma_{EFGH}\eta)\Gamma^{\kappa}\Gamma_A\Gamma^{EFGH}\Gamma_B(\psi,\<\phi^A,\phi^B\>)
\eea
We now see that, by choosing 
\bea
6\beta - \alpha\gamma &=& 0\cr
6\beta + \alpha\varepsilon &=& 0\label{relations}
\eea
(The first equation is required from coupling the scalar to the gauge field) we get closure on the fermionic equation of motion
\bea
\Gamma^{\mu}D_{\mu}\psi + \frac{\alpha}{2}\Gamma_{AB}(\psi,\<\phi^A,\phi^B\>) = 0.
\eea
That is, on-shell,
\bea
[\delta_{\eta},\delta_{\epsilon}]\psi &=& -2\gamma i(\bar{\epsilon}\Gamma_{\kappa}\eta)D^{\kappa}\psi + (\psi,\Lambda).
\eea
The gauge parameter $\Lambda = -i\alpha\varepsilon(\bar{\epsilon}\Gamma_{AB}\eta)\<\phi^A,\phi^B\>$ here, coincides with earlier expressions for the gauge parameter in Eqs (\ref{gauge1}) and (\ref{gauge2}) by means of Eq (\ref{relations}). 

The gauge invariance of the fermionic equation follows from repeated use of Jacobi identities as follows,
\bea
\delta (\psi,\<\phi^A,\phi^B\>) &=& ((\psi,\Lambda),\<\phi^A,\phi^B\>) \cr
&&+(\psi,\<(\phi^A,\Lambda),\phi^B\>) + (\psi,\<\phi^A,(\phi^B,\Lambda)\>)\cr
&=& ((\psi,\Lambda),\<\phi^A,\phi^B\>) + \varepsilon(\psi,[\Lambda,\<\phi^A,\phi^B\>])\cr
&=& ((\psi,\<\phi^A,\phi^B\>),\Lambda).
\eea

We can make a supersymmetry variation of the fermionic equation of motion and get corresponding bosonic equations of motion. Among these, we in particular find the constraint equation Eq (\ref{constraint}).

Making a field redefinition 
\bea
\phi^A &\rightarrow& \frac{1}{\sqrt{\alpha}}\phi^A\cr
\psi &\rightarrow & \frac{1}{\sqrt{\alpha}}\psi
\eea
the $\alpha$'s drop out in the supersymmetry variations,
\bea
\delta \phi^A &=& i\bar{\epsilon}\Gamma^A\psi\cr
\delta A_{\mu} &=& i\bar{\epsilon}\Gamma_{\mu}\Gamma^A\<\psi,\phi_A\>\cr
\delta \psi &=& -\varepsilon\Gamma^{\mu}\Gamma_A\epsilon D_{\mu}\phi^A - \frac{\varepsilon}{6} \Gamma_{ABC}\epsilon (\phi^A,\<\phi^B,\phi^C\>) 
\eea
and also from the equations of motion. There is no continuous parameter in this theory if the gauge field fixes the overall normalization of the action. Supersymmetry does not tell us which sign $\varepsilon = \pm 1$ to choose. We do not know what will eventually determine this sign.

It seems likely a supersymmetric action can be derived from the equations of motion one gets by making supersymmetry variation of the fermionic equation of motion, or simply by making a suitable ansatz. I have not yet worked out all the details, but it is clear that for the gauge field there is a Chern-Simons term plus the couplings to scalars and fermions via gauge covariant derivatives in their kinetic terms. Also there is a sixtic scalar interaction term of the form $(\phi^A,\<\phi^B,\phi^C\>)(\phi_A,\<\phi_B,\phi_C\>)$. This is very similar to a term that was proposed in this action in \cite{Basu:2004ed}, and in fact it become identical to it if we specify that $\varepsilon = 1$ and let $\<,\>$ be an ordinary commutator and $(,)$ and ordinary anti-commutator. More work is needed before we can tell how to realize our algebra with such multiplication operations though. At this stage it is not clear which sign of $\varepsilon$ one should choose. It is likely though that the sign of $\varepsilon$ will get determined when one has the supersymmetric action. If then $\varepsilon$ multiplies some kinetic term, demanding positive kinetic energy will fix the sign of $\varepsilon$.  

One may of course wonder how it is possible to get a parity violating Chern-Simons theory at the conformal fixed point, when the action on parallel D2-branes is parity invariant. Another issue is what the level of this Chern-Simons theory would be. These questions were originally raised in \cite{Schwarz:2004yj}, where it was also proven that no Chern-Simons theory in three dimensions with $N=8$ supersymmetry exists! Have we done something impossible then? The answer is no, since that proof relied on an ordinary Lie algebra structure of the gauge group. In this Letter we have demonstrated that one has to go outside a lie algebra in order to be able to close supersymmetry on-shell.

\section{Speculations on an infinite-dimensional realization of the algebra}
Motivated by my earlier work on the M5 brane where I have introduced fields defined on loops $C^{\mu}(s)$ in the world volume of the M5 brane, I would now like to suggest that the scalar fields are non-abelian versions of such loops, that is,
\bea
\phi^A(s)
\eea
parametrized by $s\in [0,2\pi]$. This describes a (non-abelian or matrix valued) loop in transverse space to the M2 brane, and will depend on a point $x^{\mu}$ in the M2 brane in the usual fashion, just like any ordinary local field. That is, we have the following dependence on $s$ and $x$ of our non-abelian loops: $\phi^A(s,x)$. Though we will not always display the explicit $x$-dependence.

We then suggest that the two-bracket is given by 
\bea
\<\phi,\varphi\> = \int ds \(\phi(s)\dot{\varphi}(s)-\varphi(s)\dot{\phi}(s)\)
\eea
for any non-abelian loops $\phi$ and $\varphi$. They may even be fermionic loops. A strange feature of the two-bracket is to be that it does not vanish even if the loops did not take values in a matrix algebra (that is, even if they were abelian), despite
\bea
\<\phi,\varphi\> = -\<\varphi,\phi\>.
\eea
This non-vanishing is because the two-bracket is not an ordinary commutator. 

Supersymmetry suggest the fermions are non-abelian loops $\psi(s)$ in the fermionic part of  superspace. The gauge field carries no physical degrees of freedom, and we see no reason to let this depend on $s$, so we assume that the gauge field is just an ordinary (flat) gauge field $A_{\mu}(x)$.

\section{Reduction from M2's to D2's}
We reduce by taking point-like loops (whatever that means) 
\bea
\phi^A(s) = \phi^A
\eea
except the winding loops, which can not shrink due to their topological obstruction. We take them to be minimal, that is, if $\phi^{(10)}$ is a compact (transverse) dimention with radius $R$, then we take
\bea
\phi^{(10)}(s) = Rs\label{winding}
\eea
We have no idea how these fields behave in their internal spaces. How to reduce from a field in $\phi^A(s)\in \a$, to a field $\phi^a$ that takes its values in the adjoint representation in an ordinary Lie algebra associated to the gauge group of the associated Yang-Mills theory, is to us a mystery. 

We now find that for instance a term like
\bea
\Gamma_A\int \(\psi(s)\dot{\phi}^A(s) - \phi^A(s)\dot{\psi}(s)\)
\eea
reduces to
\bea
\Gamma_{(10)} \int ds \(\psi R - Rs \dot{\psi}(s)\) = \Gamma_{(10)} R \int ds \frac{d}{ds}(s\psi(s)) = \Gamma_{(10)} R \psi
\eea
where thus 
\bea
\psi(s) = \psi
\eea
is taken to be a point-like loopino. Hence we find that the supersymmetry variation
\bea
\delta A_{\mu} = i\bar{\epsilon}\Gamma_{\mu}\Gamma_A\<\psi,\phi^A\>
\eea
reduces to 
\bea
\delta A_{\mu} = i\bar{\epsilon}\Gamma_{\mu}\Gamma_{(10)}\psi 
\eea
of super Yang-Mills, where we rescale $\phi^a\rightarrow R\phi^{a}$ and $\psi\rightarrow R\psi$. Likewise, 
\bea
\delta\psi = \Gamma^{\mu}\Gamma_A D_{\mu}\phi^A+\frac{1}{6}\Gamma_{ABC}\epsilon(\phi^A,\<\phi^B,\phi^C\>)
\eea
reduces to 
\bea
R\delta\psi = R\Gamma^{\mu}\Gamma_{(10)} D_{\mu}\phi^{(10)} + R\Gamma^{\mu}\Gamma_A D_{\mu}\phi^A + \frac{R}{2}\Gamma_{ab}\Gamma_{(10)}\epsilon[\phi^a,\phi^b].
\eea
This is almost the supersymmetry variation one has in Yang-Mills, apart from the field strength term. But this term we can also get by including point-like loops in addition to the winding loops. It seems likely one should really sum over all loops in some kind of path-integral. Then one should try to understand why only the minimal loops give the important contributions. Of course this presentation is rather vague. At the time of this writing, no known way to reduce from membrane theory to Yang-Mills theory is known. The purpose of this presentation is to give some idea of how this could work, not to give a final answer to that question. 

Since one wants to have translational invariance\footnote{We study rigid translations of the loops, so $v^A 1$ does not depend on $s$.}
\bea
\phi^A(s) \rightarrow \phi^A(s) + v^A 1
\eea
it has been suggested the three-bracket should be a Nambu bracket \cite{Berman:2006eu}, \cite{Copland:2007by}, \cite{Basu:2004ed}. That would work nicely since $1$ commutes with everything and translational invariance would get restored, though these papers are concerned with a reduction of the M2 brane theory to a fuzzy three sphere \cite{Ramgoolam:2001zx}. 

If the field is really a loop that winds the compact dimension, then $\<\phi,\varphi\>$ 
is invariant under translations. But for a different reason than that $1$ commutes with everything. It is invariant under translations because
\bea
\int ds \dot{\phi}(s) = 0
\eea
when integrated along a closed loop.

\vskip 0.5truecm

\noindent{\sl{Acknowledgements}}:

\noindent{I would like to thank N. Lambert and N. Copland for instructive discussions, and U. Gran for checking gamma matrix identities using the computer program GAMMA.}

\vskip 0.5truecm

\newpage
\appendix
\section{Gamma matrix identities}
In $11$ dimensions, the dimension of the spinor is $2^5 = 32$. Since $2^{11} = 2.(2^5.2^5)$, we get two copies of $32\times 32$ matrices if we sum from $0$ to $11$. So we should just sum from $0$ to $5$ in 
\bea
\epsilon\bar{\eta} = \sum_{n=0}^{5} c_n\bar{\eta}\Gamma_{M_1...M_n}\epsilon\Gamma^{M_1...M_n}
\eea
Making the split $11=3+8$, we find a basis
\bea
1,\Gamma^A,\Gamma^{AB},...\cr
\Gamma_{\mu},\Gamma_{\mu}\Gamma_A,\Gamma_{\mu}\Gamma_{AB},...
\eea
which constitute $2^8.8 = 2^{10}$ linearly independent matrices, as suitable for a basis of $32\times 32$ matrices.

If $\epsilon$ is Weyl,
\bea
\Gamma\epsilon = \epsilon
\eea
then we have just even numbers of $\Gamma^A$ matrices,
\bea
\epsilon\bar{\eta} &=& \sum_{p}c_p\bar{\eta}\Gamma_{A_1A_2...A_p}\epsilon\Gamma^{A_1A_2...A_p}\cr
&& + \sum_{p}d_p\bar{\eta}\Gamma_{\mu}\Gamma_{A_1A_2...A_p}\epsilon\Gamma_{\mu}\Gamma^{A_1A_2...A_p}
\eea
We may let $p=0,1,...,4$ if we also include $\Gamma$ since
\bea
\Gamma_{A_1...A_p} = \frac{1}{(8-p)!}\epsilon_{A_1...A_pA_{p+1}...A_8}\Gamma^{A_{p+1}...A_8}\Gamma
\eea

We find that
\bea
\bar{\epsilon}\Gamma_{A_1A_2...A_p}\eta - \bar{\eta}\Gamma_{A_1A_2...A_p}\epsilon &=& \bar{\epsilon}\Gamma_{A_1A_2...A_p}\eta \(1-(-1)^{\frac{p(p+1)}{2}}\)\cr
\bar{\epsilon}\Gamma_{\mu}\Gamma_{A_1A_2...A_p}\eta - \bar{\eta}\Gamma_{\mu}\Gamma_{A_1A_2...A_p}\epsilon &=& \bar{\epsilon}\Gamma_{\mu}\Gamma_{A_1A_2...A_p}\eta \(1+(-1)^{\frac{p(p-1)}{2}}\)
\eea
which, together with the Weyl property, means that the only non-vanishing combinations are 
\bea
&\bar{\epsilon}\Gamma^{AB}\eta - \bar{\eta}\Gamma^{AB}\epsilon,&\cr
&\bar{\epsilon}\Gamma_{\mu}\eta - \bar{\eta}\Gamma_{\mu}\epsilon,&\cr
&\bar{\epsilon}\Gamma_{\mu}\Gamma_{ABCD}\eta - \bar{\eta}\Gamma_{\mu}\Gamma_{ABCD}\eta.&
\eea
and we get
\bea
\epsilon\bar{\eta}-\eta\bar{\epsilon} &=& a\bar{\epsilon}\Gamma_{AB}\eta\Gamma^{AB}(1+\Gamma)\cr
&&+b\epsilon\Gamma_{\mu}\eta\Gamma^{\mu}(1+\Gamma)\cr
&&+c\bar{\epsilon}\Gamma_{ABCD}\eta\Gamma^{\mu}\Gamma^{ABCD}
\eea

We also find that
\bea
\bar{\psi}\Gamma_{A_1...A_p}\psi &\sim & 1+(-1)^{\frac{p(p+1)}{2}}\cr
\bar{\psi}\Gamma_{\mu}\Gamma_{A_1...A_p}\psi &\sim & 1-(-1)^{\frac{p(p-1)}{2}}
\eea
so, if $\psi$ is anti-Weyl, the only non-vanishing bilinears are
\bea
&\bar{\psi}\psi,&\cr
&\bar{\psi}\Gamma_{ABCD}\psi,&\cr
&\bar{\psi}\Gamma_{\mu}\Gamma_{AB}\psi.&
\eea
For $\psi$ taking values in an algebra $\a$, we find an additional minus sign (due to the antisymmetry of the bracket $\<.,.\>$) and so the above will instead be the {\sl{vanishing}} fermionic bilinears. 

For the numerical factors, we need
\bea
\delta_{A_1A_2...A_p}^{A_1A_2...A_p} &=& \(^8_p\)\cr
\Gamma_{A_1...A_p}\Gamma^{A_1...A_p} &=& (-1)^{\frac{(p-1)p}{2}}\frac{8!}{(8-p)!}
\eea
from which we get
\bea
\tr\(\Gamma_{A_1A_2...}\Gamma^{A_1'A_2'...}\)=(-1)^{\frac{(p-1)p}{2}}p!\(\tr \bf{1}\)\delta_{A_1A_2...}^{A_1'A_2'...}
\eea
Here $\tr {\bf{1}} = 2^5 = 32$. We now get the Fierz identity
\bea
\epsilon\bar{\eta}-\eta\bar{\epsilon} &=& \frac{1}{16}\Bigg[\(2\bar{\epsilon}\Gamma_{\mu}\eta\Gamma^{\mu}-\bar{\epsilon}\Gamma_{AB}\eta\Gamma^{AB}\)\frac{1}{2}\(1+\Gamma\)\cr
&&-\frac{1}{24}\bar{\epsilon}\Gamma_{\mu}\Gamma_{ABCD}\eta \Gamma^{\mu}\Gamma^{ABCD}\Bigg]
\eea

If we let $\Gamma^{M_k}$ be some completely antisymmetrized product of gammas, then 
\bea
\bar{\epsilon}\Gamma^{M_1}...\Gamma^{M_n}\eta - \bar{\eta} \Gamma^{M_1}...\Gamma^{M_n}{\epsilon} &=& \bar{\epsilon}\(\Gamma^{M_1}...\Gamma^{M_n} - e \Gamma^{M_n}...\Gamma^{M_1}\)\eta
\eea
where $e$ is the product of the signs relating $\Gamma^{M_k}$ to its transponate. For $M_k$ being single indices, we have
\bea
{\Gamma^{M_1...M_k}}^T = e_k C\Gamma^{M_1...M_k}C^{-1}
\eea
with sign factor 
\bea
e_k = (-1)^{\frac{k(k+1)}{2}}
\eea
So for instance, we get
\bea
\bar{\epsilon}\Gamma^A\Gamma_{CDE}\eta - \bar{\eta}\Gamma^A\Gamma_{CDE}\epsilon &=& \bar{\epsilon}\{\Gamma^A,\Gamma_{CDE}\}\eta\cr
&=& 6\delta^A_{[C}\bar{\epsilon}\Gamma_{DE]}\eta\cr
\bar{\epsilon}\Gamma_{\mu}\Gamma_A\Gamma_{CDE}\eta - \bar{\eta}\Gamma_{\mu}\Gamma_A\Gamma_{CDE}\epsilon &=& \bar{\epsilon}\Gamma_{\mu}[\Gamma_A,\Gamma_{CDE}]\eta\cr
&=& 2\bar{\epsilon}\Gamma_{\mu}\Gamma_{ACDE}\eta
\eea
 
Some gamma matrix identities,
\bea
\{\Gamma^A,\Gamma_{CDE}\} &=& 6\delta^A_{[C}\Gamma_{DE]}\cr
\Gamma_{ABC}\Gamma^{EF}\Gamma^C &=& 2\Gamma^{EF}\Gamma_{AB}-8\delta^{[E}_A\Gamma^{F]}\Gamma_B - 32\delta_{[A}^{[E}\Gamma_{B]}{}^{F]}\cr
\Gamma_{A}\Gamma^{EF}\Gamma_B &=& \Gamma^{EF}\Gamma_{AB}+4\delta^{[E}_A\Gamma^{F]}\Gamma_B\cr
\Gamma_{ABC}\Gamma^{EFGH}\Gamma^C &=& -2\Gamma_{[A}\Gamma^{EFGH}\Gamma_{B]}\cr
\Gamma_A\Gamma^{EFGH}\Gamma^A &=& 0 \label{gammas}
\eea

\end{document}